\documentstyle[sprocl,epsf]{article}
\def\beq{\begin{equation}}   \def\eeq{\end{equation}}

\begin{document}

\title{ EXACT RESULTS IN GAUGE THEORIES: PUTTING 
SUPERSYMMETRY TO WORK\\
\vspace{0.2cm}
The  1999 Sakurai Prize Lecture$\,$\footnote{ 
 Based on the talk at the Centennial Meeting of The American Physical
Society, March 20-26,  Atlanta, GA, on the occasion of receiving
the 1999 Sakurai Prize for Theoretical Particle Physics. Report-no.
TPI-MINN-99/20-T, UMN-TH-1759/99.}}

\author{M. SHIFMAN}

\address{ Theoretical Physics Institute, University of Minnesota, 
Minneapolis, MN 55455}

\maketitle\abstracts{
Powerful methods based on supersymmetry allow one to find exact 
solutions to certain problems in  strong coupling gauge theories.
The  inception of some of these methods (holomorphy in the gauge 
coupling  and other chiral parameters, in conjunction with  
instanton 
calculations) dates back to the 1980's. I describe the early exact 
results -- the calculation of the $\beta$ function and the gluino
condensate~--  and  their impact on the subsequent developments. 
A brief discussion of  the recent  breakthrough discoveries where 
these results play a role is given.}

\section*{Preamble}

When the question of this talk arose Arkady Vainshtein, Valya 
Zakharov and I had to decide how to split the contents into three parts. 
The division that seemed natural was that I got the part covering the
analytic properties of  supersymmetric  gauge theories, the exact 
results
following from these properties, and the  implications for
nonperturbative
gauge dynamics.  Before  delving into  the depths of
this fascinating topic let me make a few historic remarks.

I vaguely remember the seminar given by Yuri Golfand$\,$\cite{YG} 
at the  end of 1970 or the beginning of 1971 entitled something like
``Extensions  of the Poincar\'{e} algebra  by bispinor generators".
In those days I knew too little about high energy physics to 
understand the contents of the talk, let alone the novelty of the idea 
of supersymmetry (SUSY) and its potential.  My experience was 
limited, as I 
started studying  theoretical high energy physics  only a year before, 
although this was my fifth year at the Moscow  Institute for Physics 
and 
Technology. Before  that I was specializing in the dynamics of gas 
flows. The choice of the subject was not mine, I was just assigned to a 
group  of students whose major was gas dynamics and whose final 
destination was one of many classified laboratories doing research 
for 
the military.  I made several attempts to switch to more fundamental 
disciplines, but this was not allowed. This was a common practice, 
our choices were always made for us by somebody 
else. I managed to get into another group of students, specializing in 
high energy physics, only around 1970, with the help of 
V.B.~Berestetskii, who became, for a short time,  my first physics 
adviser. 

 I remember very well, however, the paper$\,$\cite{1} of Volkov and
Akulov ``Is  the neutrino a Goldstone particle?". It 
appeared in 1973,
when I had just  started working on my PhD. Now we would say that the
work was  devoted to the issue of the nonlinear realization  of
supersymmetry  and the occurrence of a massless Goldstino. It 
produced
an impression  on me. I started  pestering  colleagues,  who were a
couple of years  older, with whom I  shared the attic of the old 
mansion
occupied by  the Theory Department of the Institute of Theoretical 
and 
Experimental Physics (ITEP), the dovecote as we called it, with
questions of whether the work of Volkov and Akulov, and the idea in 
general, were worth studying. The unanimous  conclusion of the 
``elders" was negative.  In retrospect, this was evidently  the wrong
recommendation, and I feel sorry that I took it for granted. Well, in 
retrospect everything seems pretty obvious; it is much harder to 
recognize the future  potential of ideas at their birth, 
especially if one is a beginner  in the field. Sometimes
I think that even the pioneers of supersymmetry -- Scherk, Ramond, 
Golfand, Volkov, Wess,  Zumino, and others -- could not  foresee in 
the  early 1970's  that they had been opening to us the gates of the 
superworld,  which 
would become one of the most important components of our 
understanding  of Nature, a component that will stay with us forever.

It should be added that this was the  time of  the triumph of 
non-Abelian gauge theories, when quantum  chromodynamics (QCD), 
{\em the}  theory  of
hadrons, was born. This was  a new unexplored area, 
closely related to experiment,  which was rapidly developing. Valya 
Zakharov and Arkady Vainshtein got me involved in QCD.  This was 
the  type of physics I liked, and  I  submerged in it so deeply that  
what  was  happening outside was of no concern to me. Thus, the 
first decade of  supersymmetric theories, when  some of the most 
beautiful results were  obtained (e.g.  vanishing
of the vacuum energy,  nonrenormalization 
theorems,$\,$\cite{NRT} and so on) slipped by.

When I look back, I recollect these days with a nostalgic feeling.
Theory and experiment went side by side. Experimental puzzles and 
unanswered questions  that had been accumulating over the 
previous  
decade  were unfolding one after another, the solutions being
provided by the most  fundamental  theory of the day.  The game 
was 
fascinating --  we felt  that all  appropriate pieces of the riddle were 
finally there,  for the first  time  in many years.  Bits and pieces of 
knowledge started being melded 
in  a big picture. Theoretical developments, in turn, were prompting 
what was to be done next in experiment. There was a live dialogue 
between theorists and experimentalists, at the end of the day 
theoretical calculations would produce a number which could be 
tested immediately or, at least,  in the near future.   Will this 
time ever repeat itself? 

\begin{figure}   
\epsfysize=6cm
\centerline{\epsfbox{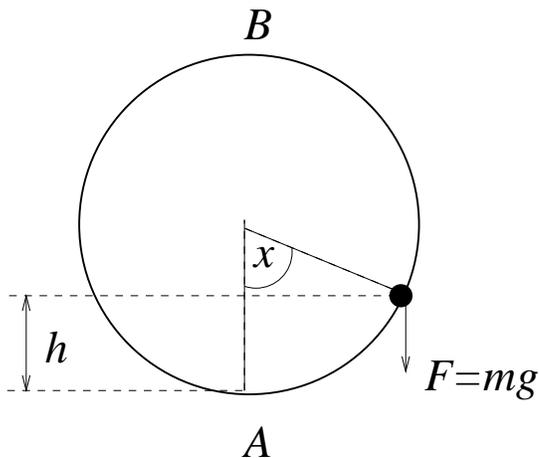}}
 \caption{Quantum mechanics of a particle $(\bullet )$ on a 
one-dimensional
topologically nontrivial manifold, the circle.}
\end{figure}

It was not until 1981 when my attention was attracted in earnest to 
supersymmetric theories. The major role in this turn of events 
belongs to Witten's paper$\,$\cite{3} {\em Dynamical Breaking of 
Supersymmetry}. It  discusses, in general terms, why 
supersymmetry could be  instrumental in the solution of the hierarchy 
problem, and why  instantons could play a distinguished role in 
supersymmetric  theories. By that time  Zakharov, 
Vainshtein and I  had been studying  instanton effects in QCD 
for several years. Instantons, discovered$\,$\cite{BPST} in 1975, 
revealed 
one of the  most profound features of  non-Abelian gauge theories 
-- 
the  existence of a nontrivial topology in the space of 
fields.$\,$\cite{4}
One of infinitely many coordinates describing the space of fields 
has the topology of the circle. To get an idea of the underlying 
physics,
one can consider a simple analog problem from quantum mechanics.
Consider a particle  in the gravitational field confined to a circle 
oriented
vertically (Fig. 1). The potential energy of the particle is
\beq
V= gh = gR\left(1-\cos x \right)\, .
\eeq

\begin{figure}   
\epsfysize=6cm
\centerline{\epsfbox{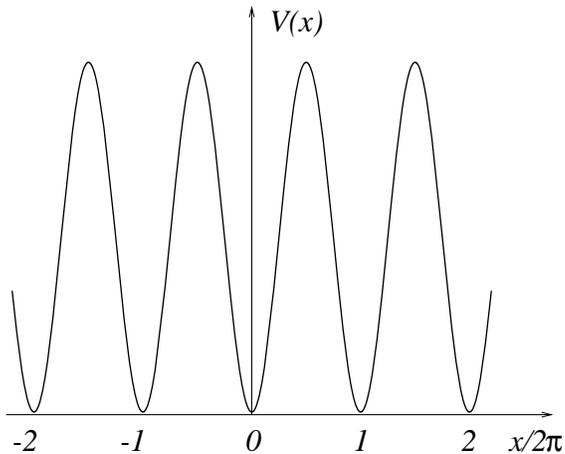}}
 \caption{If we unwind the circle of Fig. 1 onto a line we get a 
periodic 
potential.}
\end{figure}

If the kinetic energy of the particle is small enough, classically it 
oscillates near the bottom (point $A$). The fact that the circle is 
closed 
at the top (point $B$) plays no role. Only at high energies does the
classical particle  feel that it lives on the circle, since its trajectory
can wind around. 
Quantum-mechanically the possibility
of winding drastically affects even  the ground (lowest-energy)
state of the
system. The
particle can tunnel under the potential  barrier near 
the top, and return to the very same point $A$ ``from the other side". 
To solve the problem quantum-mechanically we must cut the circle 
and map it (many times) onto a line (Fig. 2). All  wave functions have
the Bloch form; in particular, the ground state wave function is
\beq
\Psi (x) = \sum_{n=-\infty}^{\infty} e^{in\vartheta} \psi_n (x)\, ,
\eeq
where $\psi_n (x)$ is the wave function of the $n$-th ``prevacuum", 
corresponding to oscillations near the point $n$ in Fig. 2, and 
$\vartheta$ is the vacuum angle, an analog of the Bloch 
quasimomentum. In QCD the circle  variable (analogous to 
the angle $x$ in Figs. 1, 2) is a composite field
built  from the gluon four-potential,
\beq
{\cal K} =\frac{g^2}{32\pi^2}\,  \int K_0 (x) d^3 x 
\eeq
where
\beq
K_\mu = 
2\varepsilon_{\mu\nu\alpha\beta}\left(
A_\nu^a\partial_\alpha A_\beta^a + \frac{g}{3}f^{abc} A_\nu^a 
A_\alpha^b A_\beta^c
\right)
\eeq
is the so-called Chern-Simons current. Winding around the circle $n$ 
times corresponds to shifting ${\cal K}$ by $n$ units. 
        
A remarkable phenomenon occurs when the gluon fields are coupled 
to massless fermions (quarks). Each tunneling in ${\cal K}$
(i.e. ${\cal K}\, \to\,  {\cal K}+1$) is accompanied,  by  necessity,
with the production of a pair of quarks of each flavor with chirality
violation.$\,$\cite{5} This can never happen at any finite  order of
perturbation theory, where  the chirality is conserved. The
instanton-induced quark vertex was found by  't Hooft, it goes under
the name of the 't Hooft interaction. 

Although  instantons in QCD were
instrumental in establishing the non-trivial  vacuum structure, the
existence of the vacuum angle $\vartheta$, and in  the qualitative
solution of the $\eta '$ problem,$\,$\cite{6} all attempts to exploit  
them 
for a quantitative solution of QCD seemed  fruitless.$\,$\footnote{
It would be more exact to say that this was our feeling
in the early 1980's. The instanton liquid models of
the QCD vacuum suggested somewhat later$\,$\cite{Shu} were
 perfected in the last decade to the extent
 that they reportedly
capture all basic regularities acting in the low-energy
hadronic physics.} 
Any sensible calculation would drag instantons into the domain of 
large radii, where the coupling constant becomes large and 
theoretical
control is lost.  We spent a  lot of time and effort trying to identify 
uses of instantons in the theory of hadrons. The outcome was not 
very 
inspiring.  Our results were limited to a few semiquantitative 
observations,$\,$\cite{7}  and one curious calculation$\,$\cite{ABC}
 which proved to be crucial in  supersymmetric theories.

The research project   which is the subject of this talk spanned many 
years,
approximately from 1981 till 1991.  When I say ``we" implying 
the authors of the project, I should be more definite.
From 1981 till 1985 our group included Novikov, Vainshtein, 
Zakharov,
and myself (as friends joked,   ``the gang of four"). 
In one crucial link we joined our forces with Misha Voloshin.
Beginning in 1986 I worked on this project  with Arkady
Vainshtein.

\section*{The puzzle of the 't Hooft interaction 
 in supersymmetric\\
 gluodynamics}

When we began thinking of   supersymmetric gauge theories in
1981,  the question of the instanton effects surfaced immediately. In 
supersymmetric gauge theories, massless fermions (gauginos, or 
gluinos 
--  I will use these terms indiscriminately), are
the superpartners of gauge bosons, which one cannot switch  off. 
A gaugino  interaction of the 't~Hooft type  is generated by 
instantons. There was no doubt in that. At the same time, there
was no doubt that  this interaction  was forbidden by 
supersymmetry, 
which  requires
every fermion vertex to be accompanied by a bosonic partner. In the 
theory with massless fermions, there are no purely bosonic type 
instanton transitions. In other words, there is no boson counterpart 
to  the  't~Hooft interaction. 

Surprisingly, this problem was not considered   in the literature
at that time.  The paradox was clear-cut, the effect was qualitative, 
and 
yet  there was complete silence in the literature  regarding this issue.
We  talked to experts,  carried out a  literature   search, 
and found next-to-nothing. 
In general, most studies of supersymmetry were limited to 
perturbative aspects. There was little effort to marry
nonperturbative gauge dynamics with supersymmetry, although
non-Abelian supersymmetric theories were known$\,$\cite{Sergio} 
since 1974.
Witten's paper,$\,$\cite{EW}  where his famous index was 
introduced, 
could  be,  perhaps, viewed as the first work where the topic of
nonperturbative  gauge dynamics was addressed in earnest. 
Then, there was a paper$\,$\cite{AHW} by  Affleck, Harvey and 
Witten
which dealt with the instanton-induced effective  superpotentials in 
three-dimensional field theories.  This work was very elegant, but -- 
alas -- it was of little help. It did not address the problem  that 
preoccupied us. 
It should be added that we were  deeply involved, for quite a time, 
with  the instanton puzzle  when these papers appeared. 

In the beginning,  the theory we mostly worked with was the 
simplest
non-Abelian supersymmetric model in four dimensions, 
supersymmetric gluodyna\-mics,$\,$\cite{Sergio}
\begin{eqnarray}
{\cal L} \!\!&=&\!\! -\frac{1}{4g^2} G_{\mu\nu}^a G_{\mu\nu}^a 
+ \frac{\vartheta}{32\pi^2} G_{\mu\nu}^a \tilde G_{\mu\nu}^a 
+\frac{i}{g^2}
\lambda^{a \alpha} 
{\cal 
D}_{\alpha\dot\beta}\bar\lambda^{a\dot\beta}\nonumber\\[0.2cm]
\!\!&=&\!\!
\frac{1}{4} \, \left( \frac{1}{g^2}- i\,\frac{\vartheta}{8\pi^2}\right)
 \int\!{\rm d}^2\theta \,\mbox{Tr}\, W^2 + \, 
\mbox{H.c.} \, ,
\label{susyym}
\end{eqnarray}
where the second line is given in 
the superfield notation, $G_{\mu\nu}^a$ is the gluon field strength
tensor,   $\vartheta$ is the  vacuum angle,  and  $\lambda^a$
is the  gluino field in the  Weyl representation. 
Note that the (inverse) coupling constant gets complexified
in supersymmetric theories, see the second line in Eq. 
(\ref{susyym}).
This circumstance has far reaching consequences, as  will be seen 
shortly.

\begin{figure}   
\epsfxsize=5cm
\centerline{\epsfbox{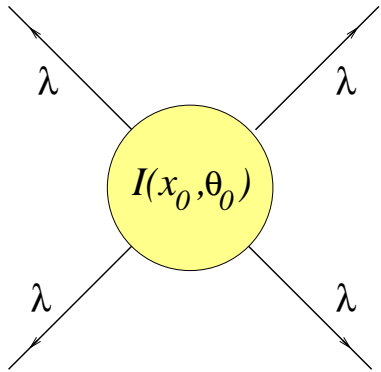}}
 \caption{The instanton-generated 't~Hooft  interaction in SU(2) 
supersymmetric
gluodynamics.  $x_0$ and $\theta_0$ are the (super)coordinates of
the instanton center.}
\end{figure}

If the gauge group is SU(2), there are four gluino  zero modes in the 
instanton background field; consequently, the 't~Hooft  vertex 
generated  by the  instanton represents a four-fermion interaction 
of the type $\lambda^4$.  The anti-instanton gives rise to 
$\bar\lambda^4$ (Fig. 3).

At the classical level, the Lagrangian~(\ref{susyym})
is invariant under  chiral U(1) rotations, $\lambda\to
\lambda\exp (-i\alpha)$. This is a valid symmetry in perturbation 
theory.
In the full theory it is absent, however. 
The instantons reveal the anomalous nature of
the  chiral U(1) through
the 't~Hooft  interaction which violates  U(1) charge 
conservation,  see  Fig.~3. Nonetheless, a
discrete subgroup $Z_4$ survives (in the case of
SU($N$) the discrete chiral invariance is $Z_{2N}$).

Given an instanton of size $\rho$, it was not difficult to 
calculate  the  coefficient of the  four-fermion interaction in 
order to check that it did  not vanish for accidental reasons. 
Sure enough,  it did not. Paradoxically, the failure of our early attempt to 
supersymmetrize the 't~Hooft interaction was because our focus on  
supersymmetry was too  narrow. Certainly, we understood that the 
family of the instanton solutions possessed a wider symmetry, 
superconformal. The superconformal group includes, in particular, the 
scale transformations which change the instanton size. Since our task 
was checking supersymmetric Ward identities we believed, however, 
that the instanton 
size $\rho$ could be kept fixed.

For over a year this problem was a constant nightmare. 
At a certain point we became so desperate that  we started 
to suspect that  SUSY 
was incompatible with nonperturbative effects, an absolutely
crazy idea.  The first relief from this agony came 
when we considered$\,$\cite{fup} the Higgsed version of the model
(\ref{susyym}). In the SU(2) model we added a Higgs sector, with 
relatively heavy  (physical) Higgs fields. The Higgs sector generated a 
mass for  the gluons  and gluinos. The four gluino zero modes I 
mentioned above have  a very  transparent geometrical meaning. 
Two  are related to  the supersymmetry
of  the model, and two correspond
to the (classical) superconformal symmetry of 
the Lagrangian (\ref{susyym}). The Higgs mass eliminated the  
superconformal invariance, and   gone with it were  the 
superconformal  zero 
modes. The two-fermion 't~Hooft vertex generated by the
remaining zero modes turned out to be a total derivative, $\partial^2
(\lambda  \lambda )$.  The corresponding contribution in the action
vanishes, and   there is no contradiction with supersymmetry.

This was a hint --  the paradox we got stuck in, was  due to our
(incorrect) presumption that one could fix the instanton size without
affecting supersymmetry.  In fact, once  one shifts in the ``fermion
direction" in the instanton moduli space, the scale transformations 
and 
the supersymmetry transformations get entangled.$\,$\cite{nsvvz} 
One 
cannot expect to obtain supersymmetric results unless the $\rho$ 
integration is done. In retrospect, the misconception
seems obvious. 

\section*{The  gluino condensate}

After we realized that, the story began to unfold very rapidly.
It was quickly  understood that the 't~Hooft vertex was not a 
good object to have chosen. We should have focused instead  on 
calculating observable amplitudes. The correlation function
\beq
\langle T\left\{ \lambda^a_\alpha (x)\lambda^{a\alpha}(x)\,
,\, \lambda^b_\beta (0)\lambda^{b\beta}(0)\right\}\rangle
\label{tpf}
\eeq
was  the most natural candidate in the SU(2) theory, given the zero
mode structure of the instanton (see Fig. 3). This understanding --
the shift towards the observable correlators and integration over 
$\rho$ -- melted the ice.
 One evening we just sat down and  did the calculation, essentially, 
on the back of an envelope. We found that: (i)  the result 
was nonvanishing, with  no visible
boson partner (this was expected), and (ii) the correlation 
function (\ref{tpf}) turned out to be an $x$-independent constant,
\beq
\langle T\left\{\mbox{Tr} \lambda^2(x)\,
,\,\, \mbox{Tr}\lambda^2(0)\right\}\rangle_{\rm inst}
= \frac{2^{10}\pi^4}{5}\,  M^6_{\rm PV}\,\frac{1}{g^4}\exp\left\{
-\frac{8\pi^2}{g^2}\right\}\,,
\label{tpf2}
\eeq
where $M_{\rm PV}$ is the Pauli-Villars cutoff parameter.  This
was  unexpected. But this was the most favorable outcome one could 
hope for: {\em the} way out.

Indeed, supersymmetry does {\em not} forbid the  correlation
function (\ref{tpf}), provided that this two-point function is spatially  
constant,
i.e. $x$ independent. The proof is quite straightforward and is 
based on three elements: (i) the supercharge 
${\bar Q}^{\dot\beta}$ acting on the vacuum state annihilates it; 
(ii) ${\bar Q}^{\dot\beta}$ anticommutes with $\lambda\lambda$; 
(iii) the derivative $\partial_{\alpha\dot\beta}(\lambda\lambda)$ is 
representable as the anticommutator of ${\bar Q}^{\dot\beta}$ and 
$\lambda^\beta G_{\beta\alpha}$. 
One differentiates Eq. (\ref{tpf}), substitutes 
$\partial_{\alpha\dot\beta}(\lambda\lambda)$ by $\{
{\bar Q}^{\dot\beta}, \lambda^\beta G_{\beta\alpha}\}$, and obtains 
zero.$\,$\footnote{By analogy with the terminology accepted
in  topological field theory  the operator $\lambda\lambda$
can be called $Q$-closed, while the operator
$\partial_{\alpha\dot\beta}(\lambda\lambda)$ is $Q$-exact.}  Thus, 
supersymmetry requires the $x$ derivative of (\ref{tpf}) to vanish.  
It does not require the vanishing  of the correlation function {\em 
per se}. A constant is okay.
                
The instanton calculation is reliable at short distances 
$|x|\ll\Lambda^{-1}$
where $\Lambda$ is the scale parameter of the theory.
Once we get a nonvanishing constant at short distances, and once
SUSY requires it to be one and the same at any distance,
we can use Eq. (\ref{tpf2}) at $|x|\to\infty$ to apply  cluster
decomposition. The latter then implies that the gluino condensate 
develops in supersymmetric gluodynamics, and that it is 
double-valued in the 
SU(2) theory,
\beq
\langle\mbox{Tr}\lambda\lambda\rangle = \pm \, 
\frac{2^{5}\pi^2}{\sqrt{5}}\, M^3_{\rm PV}\,\frac{1}{g^2}\exp\left\{
-\frac{4\pi^2}{g^2}\right\} \,.
\label{gc}
\eeq

This result was remarkable for several reasons. First of all, we were 
able to prove$\,$\cite{NSVZ1,NSVZ2} that Eq. (\ref{tpf2}) was {\em 
exact}, in the mathematical sense. 
Perturbation theory {\em per se} gives no contribution
in the correlation function (\ref{tpf2})  to any order. 
This correlator is saturated by a single
(anti)instanton -- for two or more  instantons the number of the zero 
modes
does not match. Moreover,  the (anti)instanton background field  is 
chiral, it
preserves one half of  supersymmetry. The residual supersymmetry 
is
sufficient to nullify all loop  corrections to the  instanton
configuration. There is no $g^2$ series in this problem.
The two-point function (\ref{tpf2}) is  not
renormalized, and neither is the gluino condensate. 

At one loop, the cancellation of  
 quantum corrections in the instanton  field was known 
previously.$\,$\cite{DADV} We
generalized this  assertion to all orders, putting it on par with the 
 vanishing of the vacuum energy or  the nonrenormalization
theorem for the superpotentials.$\,$\cite{NRT} 
In this way,  a number of generalized nonrenormalization theorems
was established.

The reason why such theorems  are valid in all backgrounds which 
preserve a part of  supersymmetry (usually one half), 
is the fermion-boson degeneracy, which persists in such 
backgrounds.
The possible exception is  the zero modes, which are to be treated
separately.
This is the same phenomenon that makes the energy of the ``empty" 
vacuum
vanish.  

The instanton is just a particular example of
a magic background preserving a part of SUSY. Another example is 
provided,
for instance, by  saturated domain walls -- they were discovered in 
various 
important supersymmetric models recently.$\,$\cite{DS,CS} 
The very fact of the absence of  quantum corrections 
in  magic backgrounds is universal. Details of the
proof may vary. In the instanton  problem it is so simple that I  
cannot resist the temptation to present it here.

In supersymmetric gluodynamics the
instanton center is characterized by two collective coordinates, $x_0$ 
and 
its superpartner $\theta_0$, see Fig. 3. It is important that, because 
of the selfdual (chiral) nature of the field,  there is no 
$\bar\theta_0$.  Now, consider, say, a two-loop graph in the 
instanton background (Fig. 4). This graph has two vertices; 
its contribution can be written as an integral over $d^4x d^2\theta
d^2\bar\theta$ and $d^4x' d^2\theta'd^2\bar\theta'$. 
After one integrates over the  supercoordinates 
of the second vertex and over  $d^4x d^2\theta $ (but not 
$\bar\theta$),
one is left with the integral $\int d^2\bar\theta F(\bar\theta )$. 
The function $F$ must be invariant under the simultaneous SUSY
transformations of $\bar\theta$ and the instanton 
collective coordinates. Since there is  no
$\bar\theta_0$, the only allowed solution is $F=$ Const.
If so, the integral  $\int d^2\bar\theta F(\bar\theta )=0$, {\em quod 
erat
demonstrandum}.

Certainly, this is only the skeleton of the proof. Subtleties must be 
taken care of (e.g. the absence of   infrared divergences).
You may believe me that the statement of no corrections
in Eq. (\ref{tpf2})  is clean.

The exactness of the one-instanton result for the
 correlation functions of  appropriate chiral superfields
(the operators involved  must be the lowest components of the
superfields of one and the same
chirality, and must saturate all instanton zero
modes) is a rigorous mathematical statement. Whether the calculation
of the gluino condensate 
outlined above  is physically complete  is a different story, on which I
will dwell later. Now let me only note that it opens three distinct
directions: (i) to topological field theories; (ii) to exact $\beta$ functions;
(iii) to condensates in the strong coupling regime. I will
consider these issues in turn.

\begin{figure}   
\epsfxsize=6cm
\centerline{\epsfbox{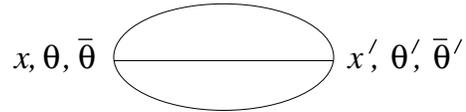}}
 \caption{A typical two-loop supergraph. The solid lines denote
the propagators of the quantum superfields in the (anti)instanton
background.}
\end{figure}

\section*{The road  to topological field theories}

The line of reasoning that led us
to the gluino condensate (see the discussion after Eq.~(\ref{tpf2}))
was a hint that the quantity we calculated was nondynamical.
Indeed, the gluino condensate was determined through arbitrarily small 
instantons.  The subsequent observation$\,$\cite{zeroSV} that
in weak coupling the gluino condensate was saturated
by {\em zero-size} instantons was  an even  stronger message.

We were discussing the  issue
over and over. Around 1986, Arkady and I did an
instructive exercise. We considered SUSY gluodynamics
in  gravitational backgrounds rather than in  Minkowski space.
Certainly, for an arbitrary background, supersymmetry is lost.
However, some
backgrounds still preserve (a part of) 
supersymmetry. An example of this type is the theory on a
four-dimensional Euclidean sphere of  radius $R$. An even  simpler
example is provided by the theory  on  a four-torus with  arbitrary
periods $L_i$. In the limit $R,L_i\to\infty$ one returns back to 
flat
space. It seemed instructive, for reasons which I will mention later, 
to  keep these dimensional parameters finite. 

In the case of the sphere, the exercise is more complex technically 
than 
the Minkowski-space calculation, while for the torus it is only 
marginally different from that in  Minkowski space. (Moreover, 
on the  torus one can find$\,$\cite{ECoh}  
$\langle\mbox{Tr}\lambda\lambda\rangle$ directly, using 
torons$\,$\cite{torons}).  We found the analog of the correlation
 function (\ref{tpf2}) in both cases  and
observed that (i) the result depends neither on $R$ nor on
$L_i$, and  (ii) the numerical coefficient in front of 
$ M^6_{\rm PV}\,g^{-4}\exp\left\{
-{8\pi^2}/{g^2}\right\}$ does depend on whether we are on the 
sphere 
or   the torus. This was  a clear indication of  the topological 
nature of  the sector of the theory under consideration. 
The finding was exciting, and we discussed the situation with 
Arkady  many times. I described  what I knew in a lecture at the
Zakopane school in May 1988. From  there
it was only one step to isolating this sector, by discarding the rest of 
the theory.
 Our mathematical
culture was not high enough, however, to make this step possible.
After I returned from Zakopane, a colleague of mine told me that
he had heard of Witten's work on  topological field 
theory.$\,$\cite{EWtop}
We went to the library to look at the preprint (it was published
in February 1988). It was
stolen. This was not unusual. Since we had essentially no access to 
photocopying
machines, interesting papers, especially Witten's,  would frequently 
disappear upon 
arrival. I do not want to say that I or any of my
colleagues were stealing from the library,  but some preprints
were just disappearing into thin air.  So, we had to wait till the 
journal
publication came. It was remarkable to see how far
Witten advanced the strategy  I described in the passage after Eq.
(\ref{tpf2}). He peeled off the dynamical contents of the 
supersymmetric
gauge theories; what remains was formulated  in a form 
preserving a
residual supersymmetry in {\em any} gravitational background. 
Topological field theories are metric-independent.  All correlation 
functions
in  topological field theories
are treatable in the same manner as we treated  the $\lambda^2$ 
two-point
function, see  Eq. (\ref{tpf}).  It is most remarkable that  
topological
field theory became a powerful tool for solving some long-standing
mathematical problems which were apparently of paramount 
importance for mathematicians.$\,$\footnote{Later I learned that a 
special 
topological field theory was suggested$\,$\cite{AS} by Albert 
Schwarz as 
early as in
1978.} 

\section*{Gluino condensate and  spontaneous breaking of 
supersymmetry}

The assertion that the correlation functions of the lowest components
of the superfields of one and the same chirality (all chiral or all 
antichiral) are coordinate-independent is valid not only for the 
gluino
operators $\mbox{Tr}\, \lambda^2$. This theorem is general  and 
is applicable to any operator. We   did not take advantage of this 
circumstance. 
It 
was  Rossi and Veneziano who initiated$\,$\cite{RV}  a 
systematic  search  for correlation functions of the type (\ref{tpf}) which 
are 
saturated by one instanton, in various theories with matter.  In 
practice, the search is quite an easy task since the analysis 
essentially reduces 
to a dimensional counting (the dimension of the appropriate 
correlation function must match the first coefficient of the
$\beta$ function)  and keeping the balance of
the zero modes. This line of research culminated in the very 
beginning of 1984  when the SU(5) model with  $M$ quintets $V$ 
and  $M$ antidecuplets $X$ was considered.$\,$\cite{MV} For
instance, for $M=1$  the appropriate correlation function is
\beq
\Pi (x,y) = \langle T\,[\mbox{Tr} \lambda^2 (x) , 
\mbox{Tr}\lambda^2 
(y),
{\cal S} (0)  ]\,\rangle\,, \qquad {\cal S}  = XXVX\lambda^2\, .
\label{mfcf}
\eeq
The color indices are contracted in ${\cal S} $ in a self-evident way, 
namely, 
$
{\cal S}  =
\epsilon^{\alpha\beta\gamma\delta\rho} 
X_{\alpha\beta}X_{\gamma\delta} \, (V^\kappa 
X_{\kappa\chi}\lambda^\chi_\psi\lambda^\psi_\rho )$
(the Lorentz indices of the gluino fields are suppressed). 
All  operators in the correlation function $\Pi (x,y)$
are the lowest components of  chiral superfields.  The 
one-instanton contribution does  not vanish
and does produce a constant times $\Lambda^{13}$.
(The number 13 looks odd; in fact, this
 is  the first coefficient of the $\beta$
function in the model at hand;  $\Lambda^{13}$ matches 
the dimension of $\Pi  (x,y)$.)
If $x,y\ll \Lambda^{-1}$ one expects that the one-instanton 
contribution saturates $\Pi (x,y)$, so that the constant obtained in 
this  way is reliable. If so, one can pass to the limit $x,y\to\infty$ 
and  use the property of clusterization at large $x,y$ to
prove  that the gluino condensate develops, $\langle \mbox{Tr}\,
\lambda\lambda\rangle \neq 0$. The solution  with 
$\langle \mbox{Tr}\lambda\lambda\rangle = 0$ and
$\langle{\cal S} \rangle 
\to\infty$ is ruled out due to the absence of  flat directions.  Since 
the
superpotential   is  absent in this model, the gluino condensate is the 
order
parameter for  SUSY breaking. One concludes that supersymmetry is 
spontaneously
broken.$\,$\cite{MV} In fact, this was the first {\em direct} 
demonstration
that nonperturbative effects in the gauge theories 
in four dimensions can lift the 
classical
supersymmetric vacua resulting in the dynamical breaking of
supersymmetry. Later on this technique was overshadowed by the
effective Lagrangian approach elaborated$\,$\cite{ADS1} by Affleck, 
Dine 
and
Seiberg (ADS). In many instances the latter is indeed
more ``user-friendly," since   it allows one to easily  trace the 
response 
of 
the theory to the continuous deformations of  parameters, starting 
from 
the weak coupling Higgs regime. The condensate-based analysis 
remains 
useful in the strong-coupling regime. There is one unsolved mystery 
associated with this analysis, which I will return to  at the end of the 
talk. 
The relation between the two approaches seems pretty obvious 
now~-- 
in 
the weak coupling they are totally equivalent. Apparently, this was 
not 
so evident  then.  I remember that shortly after the ADS papers, I 
spent 
a month at CERN in Geneva. This was my first serious exposure to the
 Western world,
I was sort of depressed by the contrast between what I saw around
and my every-day experiences in Moscow, so I decided that the best 
thing to do was not to venture outside CERN at all.  I spent 
the  entire
month confined  in the offices of Daniele Amati and Gabriele 
Veneziano.
We had endless discussions of how the transition from the
weak coupling Higgs regime to the strong coupling regime could 
occur. In the end, I left with our understandings  still far apart.
One of the conjectures was especially close to the hearts of Daniele 
and Gabriele; I did not like it then, and appreciated it only a 
decade later. The equivalence between the condensate-based
program$\,$\cite{NSVZ2,RV}  and
the ADS approach in the weak coupling regime  was
elucidated in a dedicated paper.$\,$\cite{NSVZjetpl}

\section*{Strong vs. weak coupling regime: the 
power of holomorphy}

The one-instanton contribution to the correlation functions
(\ref{tpf}), (\ref{mfcf}) does not vanish and is compatible with
supersymmetry (i.e. one gets a coordinate-independent constant).
When all coordinates  are close to each other, at 
 short distances, this is not so surprising: the result is saturated by 
small-size instantons. In  asymptotically free theories, where
the short-distance behavior is controllable, the calculation seems to 
be 
safe.
However,  the one-instanton contribution continues to be the
very same constant at large distances.
Technically this is due to the fact that at $x,y,....\to \infty$
the integration over the instanton size $\rho$ is saturated
at $\rho\to\infty$. Moreover,  there are no quantum corrections in 
the instanton
background field whose explosion could signal the failure of this 
regime.
Coherent field fluctuations of that large size do not make sense in 
conventional
confining theories. We did not feel  satisfied with our
degree of understanding
of the strong coupling calculations. Arkady and I kept trying
to get a clearer picture or, at least,  formulate a
clean  roundabout procedure
that would allow us to obtain the gluino condensate in the strong 
coupling
regime starting from the weak coupling Higgs regime, where we 
were
confident in all stages of the analysis. The guiding principle was the 
smooth transition between the weak and strong coupling domains in 
the 
theories with fundamental matter, a conjecture known in the 
literature for 
quite a time.$\,$\cite{BRFS} We debated the issue for a couple  years, 
off and on, until a strategy crystallized as  to
 how one could pin down the gluino condensate, in the 
fully controllable environment (this happened after a very 
illuminating
conversation with Gabriele Veneziano, who was visiting ITEP in late 
spring 1987). 

The basic idea was as follows.$\,$\cite{SV2} Consider, for instance, 
SU(2)  SQCD 
with one flavor. The vacuum structure in this theory  was 
found,$\,$\cite{ADS1}
in  weak coupling, by Affleck, Dine and Seiberg by 
integrating out heavy degrees of freedom and analyzing the
 effective low-energy
Lagrangian for the light degrees of freedom (moduli).  The 
Lagrangian
of the model  is
obtained by adding to Eq. (\ref{susyym}) the matter term
\beq
{\cal L}_{\rm matter}= \frac{1}{4} \int \! {\rm d}^2\theta {\rm
d}^2\bar\theta\,
\bar Q^f e^V Q_f +
\left\{ \frac{m}{4}\int\! {\rm d}^2\theta\, Q^{ f}_\alpha Q_{ f}^\alpha
+{\rm H. c.}\right\}\, ,
\label{su2lagr}
\eeq
where $ Q_{ f}^\alpha$ is a chiral superfield, $\alpha$ and $f$
are the color and subflavor indices, respectively, $\alpha , f = 1,2$.
The weak coupling regime is achieved when the matter
mass parameter  $m$ is small, $m \ll \Lambda$.  In this case the 
expectation
value of the modulus $Q^{ f}_\alpha Q_{ f}^\alpha$ is large,
\beq
\langle Q^{\alpha f}Q_{\alpha f}\rangle =\pm\, 2
m^{-1/2}\,  M^{5/2}_{\rm PV}\,\frac{1}{g^2}\exp\left\{
-\frac{4\pi^2}{g^2}\right\}\,.
\eeq
The gluons and  gluinos are heavy and are integrated out in the ADS
Lagrangian. Nevertheless,  the vacuum value of the modulus 
quoted above unambiguously determines
the gluino condensate, by virtue of 
 the Konishi relation,$\,$\cite{Konishi} namely
\beq
\langle \mbox{Tr}\lambda\lambda\rangle =
8\pi^2 m\, \langle Q^{\alpha f}Q_{\alpha f}\rangle = \pm \, 
m^{1/2} \left( {2^{4}\pi^2}\right) \, M^{5/2}_{\rm 
PV}\,\frac{1}{g^2}\exp\left\{
-\frac{4\pi^2}{g^2}\right\}\, .
\eeq

The key observation of Ref. 32 is that the square root 
dependence of the gluino condensate on the bare mass 
parameter $m$  is {\em  exact}.
It is the consequence of supersymmetry and a generalized $R$  
symmetry of the model at hand. It is possible to establish the exact 
relation
because $\mbox{Tr} \lambda\lambda$ is a chiral operator while 
$m$
is a chiral parameter; in the
modern language one says that $m$ can be promoted to an auxiliary 
chiral
superfield.   Then, $\mbox{Tr}\lambda\lambda$ can depend only on 
$m$
but not on $\bar m$.
The gluino condensate is an analytic function of $m$.
Thus, in supersymmetric theories the notion of smoothness
can be replaced for chiral quantities by an {\em exact analytic} 
dependence.
If so, by calculating  the gluino condensate at small $m$, when
the theory is weakly coupled, one can analytically continue to large 
$m$, 
i.e. 
$m \to M_{\rm PV}$, 
where the matter fields become heavy, and can be integrated out, 
thus
returning us to strongly coupled SUSY gluodynamics. And yet, we
know the gluino condensate exactly.  In this way, $\langle
\mbox{Tr}\lambda\lambda\rangle$ was found in the strong coupling
regime for all gauge groups.$\,$\cite{SV2,MOS} 

This idea -- extrapolating from weak to
strong coupling on the basis of  
holomorphy -- became a dominant theme
for  Arkady and I beginning in 1987. It
was later elevated$\,$\cite{Sei1} to new heights by Seiberg.
He considered superpotentials in  theories with  arbitrary 
gauge 
and
Yukawa couplings and established, using a similar line of reasoning,  
various
nonrenormalization theorems and a wealth of elegant exact results. 
Note that since the arguments are essentially based  only  on 
holomorphy, 
they are valid not  only
perturbatively but also nonperturbatively. The strategy of picking up 
chiral quantities with  known holomorphic behavior, calculating 
them (e.g.  through
instantons) at weak coupling, with the subsequent analytic
continuation to  strong coupling, is a standard practice now, after 
the
works of Seiberg$\,$\cite{Sei2} that shook the world.  It was quite an 
exotic endeavor back  in 1987.

\section*{The exact $\beta$ functions}

Now, let me return to the topic of  exact $\beta$ functions in 
supersymmetric
theories. In the beginning of the talk I mentioned a curious 
calculation  we
did in 1981 in (non-supersymmetric) QCD. It was 
observed$\,$\cite{ABC}
that the running of the gauge coupling $\alpha_s$, as it emerges in 
the
instanton measure, has a remarkable interpretation. As is 
well-known, the first
coefficient $b$ of the Gell-Mann--Low function can be represented 
as$\,$\cite{asfr}
(for the SU($N$) gauge group)
\beq
b = \frac{11}{3} N = \left( 4-\frac{1}{3}\right) N\,.
\eeq
Here $4N$ represents an antiscreening contribution, which in 
perturbation theory (in the physical Coulomb gauge) is associated 
with the 
Coulomb gluon  exchange
and has no imaginary part, while $-N/3$ is the ``normal" screening
contribution, the imaginary part of which is determined by
 unitarity. Within 
instanton calculus, the term $4N$ is entirely due to the zero modes. 
It  has a  geometrical meaning, and its calculation is trivial. 
The  part which is relatively hard to obtain, 
$-N/3$, comes from the nonzero modes. When we learned, from  
D'Adda and  Di~Vecchia's work,$\,$\cite{DADV} that the nonzero 
modes 
in supersymmetric theories cancel in the  instanton
measure at one loop, we immediately realized that the cancellation 
would persist to all orders, and the $\beta$ function would be 
exactly  calculable, in a technically trivial way.

In supersymmetric gluodynamics the $\beta$ function
turns out to be 
 a geometrical progression. This is seen from the  instanton
measure or, which is essentially the same, from
the gluino condensate.  Being an observable quantity,  it is certainly 
renormalization-group
invariant.  Since Eq. (\ref{gc}) is exact, an exact relation
between the ultraviolet parameter $M_{\rm
PV}$ and the bare coupling constant emerges: the explicit $M_{\rm
PV}$ dependence of the right-hand side of  Eq. (\ref{gc})  
 must be canceled by an  implicit
dependence coming from $1/g^2$.
 
In this way one gets the $\beta$ function,
\beq
\beta (\alpha) = -\frac{6\alpha^2}{2\pi}\left(
1-\frac{\alpha}{\pi}\right)^{-1}\, , \qquad \alpha =\frac{g^2}{4\pi}\,.
\label{totbetapg2}
\eeq
This is  for SU(2); for an  arbitrary gauge group
\beq
\beta (\alpha) = -\frac{3T_G\, \alpha^2}{2\pi}\left(
1-\frac{T_G\,\alpha}{2\pi}\right)^{-1}\, ,
\label{totbetapg}
\eeq
where $T_G$ is the dual Coxeter number (it is also called
 the Dynkin index). As
will be explained shortly, Eqs. (\ref{totbetapg}) and  
(\ref{totbetapg2}) 
are
exact not only  perturbatively, but nonperturbatively as well.
Our approach makes explicit that {\em all} coefficients
of the $\beta$ function have a geometric interpretation$\,$\footnote{
Even more pronounced is the geometric nature of the coefficients
 in the two-dimensional K\"{a}hler sigma models, for 
obvious reasons: these models  are geometrical.
The supersymmetric K\"{a}hler sigma models have extended
supersymmetry, ${\cal N}= 2$. Therefore, the $\beta$ function
is purely one-loop. We performed$\,$\cite{sigma} the instanton 
calculation 
of
the first coefficient for all nonexceptional
compact homogeneous symmetric K\"{a}hler manifolds.
It might seem that in theories with matter, see Eq.~(\ref{nsvzbetaf})
below, the geometrical interpretation of the second and higher 
coefficients of the $\beta$ function is lost because of the occurrence of 
the anomalous dimensions $\gamma_i$. In fact, it has been recently 
shown$\,$\cite{hori} that the running gauge coupling one obtains within  
$D$-brane engineering is compatible with
Eq.~(\ref{nsvzbetaf}). Thus, a geometric interpretation is recovered.}
 --
they count the number of the instanton
zero modes which, in turn,  is related to the number of nontrivially
realized symmetries. Indeed,
\beq
\beta (\alpha) = - \left(n_b-\frac{n_f}{2}\right)\, \frac{ 
\alpha^2}{2\pi}\left[
1-\frac{\left(n_b-n_f\right)\,\alpha}{4\pi}\right]^{-1}\, ,
\label{totbetapg3}
\eeq
where $n_b$ and $n_f$ count  the 
gluon and gluino zero modes,
respectively. In this form the result is
valid in  theories with extended supersymmetry, too. For ${\cal 
N}= 
2$,
one gets $n_b = n_f = 4T_G$, implying that the $\beta$ function is 
one-loop.
For ${\cal N}= 4$ the $\beta$ function vanishes since  $n_f = 2n_b$.

In  theories with matter, apart from the gluon 
and  gluino zero modes, one has to deal with the zero 
modes of the matter 
fermions. While the gluon/gluino $Z$ factors are related to the 
gauge coupling constant $g^2$ itself, this is not the case for the
$Z$ factors of the matter fermions.  The occurrence of the
additional $Z$ factors brings  new ingredients into the analysis,
the anomalous dimensions of the 
matter fields $\gamma_i$.
Therefore, in  theories with
matter the exact  instanton measure implies an exact relation
between the $\beta$ function and the anomalous dimensions 
$\gamma_i$,
\begin{equation}
\beta (\alpha) = -\frac{\alpha^2}{2\pi}\left[3\,T_G -\sum_i T(R_i)(1-
\gamma_i )
\right]\left(1-\frac{T_G\,\alpha}{2\pi} \right)^{-1}
\, ,
\label{nsvzbetaf}
\end{equation}
where $T(R_i)$ is the Dynkin index in the representation $R_i$,
$$
\mbox{Tr}\, (T^a T^b ) = T(R_i)\, \delta^{ab}\, ,
$$
and $T^a$ stands for the generator of the gauge group $G$; 
the latter can be arbitrary.

Equation (\ref{nsvzbetaf}), 
which is sometimes referred to as the
Novikov-Shifman-Vainshtein-Zakharov (NSVZ) $\beta$ function,
is valid for
arbitrary Yukawa interactions of  the matter fields. The Yukawa 
interactions show
up only through the  anomalous dimensions. 

The NSVZ $\beta$ function has the unique property that if one 
evolves 
the 
gauge coupling ``all the way down," till its evolution is complete and 
the 
coupling is frozen, the value of the frozen coupling  is  as 
if 
the $\beta$ function were one-loop, although, in fact, the
evolution is certainly governed by the multiloop $\beta$ function.
I will explain this point later.

In the beginning our attention was almost entirely focused on 
perturbative calculations of the $\beta$ functions. The reason is 
quite obvious --
the generalized nonrenormalization theorem
in the instanton background  we had established
is valid order by order in perturbation theory. Later 
we realized that one can
apply, additionally,
$R$ symmetries to prove that in typical models,
Eq. (\ref{nsvzbetaf}) is also valid  nonperturbatively. This aspect
is  discussed, in particular, in Ref.~40.
The fate of the assertion of  ``nonperturbative
exactness" is rather surprising: it is being
rediscovered again and again,  see e.g. fresh
publications.$\,$\cite{morariu,AHM} I hasten to add that
exceptional models, in which  the NSVZ $\beta$ function is corrected 
at the  nonperturbative level, are not rare.
The most notable one is the ${\cal N}=2$ theory
that played the key role in Seiberg and Witten's
breakthrough$\,$\cite{SEIW} in 1994. In the ${\cal N}=2$ theory the 
NSVZ $\beta$ function is 
one-loop. However, instantons generate an infinite series of 
nonperturbative 
terms, for reasons that are well understood.$\,$\cite{NS88}  The full 
$\beta$ function is rather nontrivial, it can be explicitly 
found$\,$\cite{Ritz} from the Seiberg-Witten solution. 

The master formula (\ref{nsvzbetaf})  kept us busy for several years. 
We 
derived it more than once: first  from the analysis of perturbation 
theory,$\,$\cite{little} and
then from the  consistency of the anomalies in supersymmetric
theories.$\,$\cite{SVanom} The latter topic, the consistency of the 
anomalies, has far-reaching consequences by itself.  
I will discuss it shortly. As for
implications of  the NSVZ
$\beta$ function, let me mention a few examples. An immediate 
consequence is the one-loop nature of the  $\beta$
function in ${\cal N}= 2$ extended supersymmetries and the 
vanishing 
in ${\cal N}= 4$. Of course, these facts were established long ago from 
other 
considerations. 

 More productive are the applications
where the NSVZ  $\beta$ function leads to novel results. 
For instance,  it allows one to generate finite theories even in the 
class of
${\cal N}=1$.  The simplest example  was
suggested in  Ref.~48, further developments are 
presented
in  Ref.~49. The general idea is to arrange the 
matter sector in such a way that the 
conditions $3\,T_G - \sum_i T(R_i)=0$ and $\gamma_i=0$ are met
simultaneously. For instance,
consider the SU(3) gauge model with nine triplets $Q^i$ and nine 
antitriplets $\tilde Q_i$ and the superpotential$\,$\cite{PWJM}
\begin{equation}
{\cal W}=h\left(Q^1 Q^2 Q^3 +Q^4 Q^5 Q^6 +Q^7 Q^8 Q^9 +
\tilde Q_1 \tilde Q_2 \tilde Q_3 +\tilde Q_4 \tilde Q_5 \tilde Q_6 
+\tilde 
Q_7 \tilde Q_8 \tilde Q_9\right)\,,
\end{equation}
where  contraction of the color indices by virtue of 
$\epsilon_{ijk}$
is implied. The flavor symmetry of the model ensures that there is 
only 
one $Z$ factor for all matter fields. Since the condition $3\,T_G - 
\sum_i 
T(R_i)=0$ is satisfied,  finiteness is guaranteed provided that the 
anomalous dimension $\gamma$ vanishes. At small $g$ and $h$ the 
anomalous dimension $\gamma (g, h)$ is
determined by  a simple one-loop calculation,
\begin{equation}
\gamma (g, h)=-\frac{g^2}{3\pi^2} + \frac{|h|^2}{4\pi^2}\,.
\end{equation}
This shows that the condition $\gamma (g, h)=0$ has a solution, at 
least 
for  small couplings. If the initial conditions $g_0$ and $h_0$
are chosen in such a way that $\gamma (g_0, h_0) =0$, the coupling 
constants do not run -- they  stay at $g_0\, ,\, h_0$ forever. The 
Yukawa  coupling $h$ is frozen due to the fact that the $\beta$ 
function 
for $h$ is proportional to $\gamma (g, h)$.

Straightforward extensions of the methods developed in connection with 
the NSVZ $\beta$ function yield a spectrum of exacts results
going well beyond the original range of applications. For instance,
renormalization of the soft supersymmetry breaking parameters has
been recently treated along these lines,$\,$\cite{hishi} to all orders
in the gauge coupling constant.
Among other uses, I would like to mention the determination of the
boundaries of Seiberg's conformal window.$\,$\cite{Sei2}
A\hspace{0.1cm}Êrelated issue is the determination of the conserved 
$R$
current for the theories lying in the conformal window.
We obtained (see the second paper in Ref.~40)
a unified expression which interpolates between Seiberg's current  in 
the ultraviolet and the geometric current in the infrared
conformal limit.
Furthermore,  the NSVZ $\beta$ function allows one
to exactly calculate the conformal central charges.$\,$\cite{ccc}
These are good problems; unfortunately, their discussion
 will lead us far astray.

As a curious fact, let me note that the $\beta$ function 
in supersymmetric gluodynamics  first appeared in the  form of a 
geometric progression in the paper of
Jones,$\,$\cite{Jonesr} one of many early works  devoted to the  
superanomaly problem, a topic on which   I will dwell shortly.
Both, the starting
assumption of this work  and the basic steps of derivation are 
irrelevant,
as we understand it now, and yet, paradoxically, Eq.~(\ref{totbetapg}) 
shows up. Closer to the modern understanding of
the superanomaly problem is a construction suggested
by  Clark, Piguet and Sibold.$\,$\cite{CPS}
It is very hard to  read these papers, but those who  managed to work
through them would  be rewarded by extracting  a simplified version of 
Eq.~(\ref{nsvzbetaf}),
\beq
\beta (\alpha ) = \frac{\alpha^2}{\pi} \left[1-\gamma (\alpha 
)\right]\,, 
\label{bfsqed}
\eeq
 applicable in   SUSY QED.

\section*{Three geometric anomalies and supersymmetry}

This problem  has many facets. It
lies at a junction of several deep phenomena in supersymmetric 
theories.  
To put things in the proper perspective, I should start from 1974 when 
Ferrara and
Zumino noted$\,$\cite{Sergio2} that the axial current $a_\mu$, the 
supercurrent
$S_{\mu \alpha}$ and the energy-momentum tensor 
$\theta_{\mu\nu}$
enter in one and the same supermultiplet, dubbed the supercurrent
 supermultiplet $J_{\alpha\dot\alpha}$. It is curious that 
supersymmetric gluodynamics was treated in  an Appendix to 
Ref.~54,
while the  main
body of the paper dealt with the Wess-Zumino model. 
It was proved that in  classically conformal theories
\beq
\bar D^{\dot\alpha} J_{\alpha\dot\alpha} = 0\, ,
\label{scccl}
\eeq
 while in the generic supersymmetric theories
$\bar D^{\dot\alpha} J_{\alpha\dot\alpha} =  D_{\alpha}\Phi$ where
$\Phi$ is a chiral superfield, elementary or composite.
Equation (\ref{scccl}) combines the conservation laws for the chiral 
current, supercurrent and the  energy-momentum tensor.

As is well-known, all three objects above have quantum anomalies.
It was noted, in the most explicit form by Grisaru,$\,$\cite{Grisa} 
that if
 $a_\mu$, $S_{\mu \alpha}$ and $\theta_{\mu\nu}$ form a 
supermultiplet, the
same must be valid for the corresponding anomalies.  It was 
checked$\,$\cite{Grisa}
that this is, indeed, the case at the one-loop level. 

The anomaly saga in supersymmetric gluodynamics  starts
from two loops. On the one hand, according to the Adler-Bardeen
theorem,$\,$\cite{AB} the chiral anomaly is exhausted by one loop. 
On 
the  other hand, the anomaly in the trace of the energy-momentum 
tensor
$\theta_{\,\mu}^{ \mu}$ was believed to be proportional to the 
$\beta$ 
function. It was apparently   multiloop. This discrepancy defied
supersymmetry. The contradiction was irritating, it was a dark spot 
on  the otherwise beautiful face of supersymmetry. Quite a 
significant 
effort  was invested in this problem. A couple of dozen works appeared 
in  the 
late  1970's and early 1980's suggesting various 
``solutions," to no 
avail (for a representative list of references see e.g. 
Ref.~47).
The mystery of superanomalies resisted all attempts at a 
``reasonable" solution. To give you a feeling of how desperate 
people were, in 1984 we published a paper entitled ``Anomalies are 
not 
supersymmetric. Is SUSY anomalous?".$\,$\cite{NSVZanom}  In this 
paper a no-go
theorem was established ruling out the possibility of  two chiral 
currents 
(one  of them
belonging to the supercurrent supermultiplet and another obeying 
the Adler-Bardeen theorem) that would differ by a subtraction 
constant.
This was the most popular construction on the theoretical market of 
the  day. Of course, now this theorem  has no  value  other than 
historical. 

To tell you the truth, we became obsessed with this puzzle.
The superanomaly problem was always at the back of my
mind even when I was doing something else. This went on 
for several years. I do not remember why, but in the
late spring of 1985 Arkady and I
decided to do an elementary  exercise -- find the effective action at 
two
loops in massless scalar electrodynamics. We did it in an
unconventional way, by applying the background field
technique and the Fock-Schwinger gauge for the background
photon. I remember I was wrestling with
this ``elementary  exercise" well into summer,
on vacations in P\"{a}rnu on the Baltic sea. We kept obtaining a 
nonsensical
expression until we discovered$\,$\cite{qed} a remarkable feature:
the second loop was actually infrared. It was saturated by
virtual momenta of order of the momentum of the
external photon.  This seemingly insignificant
observation opened our eyes.

In addition, approximately at the same time, we received
 two works,$\,$\cite{Gone}
which produced a very strong impression on us, in the 
technical sense. In fact, they pointed in the same
 direction.  Following  these  hints, we 
found a solution which turned out to be quite  unexpected.

In the works,$\,$\cite{Gone} the
supergraph background field technique was applied to a direct 
calculation of the  effective
action in supersymmetric gluodynamics
at two loops. The authors used the
supersymmetric  regularization 
via dimensional reduction (DR). The result for the effective action 
exhibited a  very clear
distinction between the first and all higher loops. The operator
\beq
 \int\!{\rm d}^2\theta \,\mbox{Tr}\, W^2 + \, 
\mbox{H.c.}\, ,
\label{khorop}
\eeq
which is gauge invariant with  respect to the background field, 
appears only at one loop. The second loop gives rise to a distinct 
structure, 
reducible to (\ref{khorop}) through an ``artificial" substitution which, 
at first  glance, 
seemed very suspicious to us. Indeed, in $4-\varepsilon$ dimensions,
apart from (\ref{khorop}), there exists another operator, 
${\hat\Gamma}^2$,
gauge invariant with  respect to the background field. Here 
$\Gamma$ is the
gauge connection, and the  caret means its projection onto the extra
$\varepsilon$ dimensions.$\,$\footnote{In the original publication 
the authors use two carets,
one on top of the other. Being typeset in Latex such a tower looks too 
ugly.} The two-loop supergraphs in a direct 
calculation
yield a nonchiral term
\beq
\frac{1}{\varepsilon^2} \int\!{\rm d}^2\theta {\rm
d}^2\bar\theta\, {\hat\Gamma}{\hat\Gamma}\, ,
\label{plokhp}
\eeq
which reduces to (\ref{khorop}) by virtue of the relation
$\bar\nabla^2 {\hat\Gamma}{\hat\Gamma}=-\varepsilon W^2$.
The structure (\ref{plokhp})  does not exist in four dimensions.
The operator (\ref{plokhp}) had been interpreted in$\,$\cite{Gone}
as a local counterterm leading to the distinction between the two alleged 
axial 
currents. The results of Ref.~59 taken at their face value 
-- not the interpretation suggested by the authors -- pointed in the
opposite direction: the second and higher loops in the effective action 
are in fact
an {\em infrared} effect.$\,$\cite{SVanom} In four dimensions
 Eq.~(\ref{plokhp}) should  have been converted into
\beq
\int\!{\rm d}^2\theta {\rm
d}^2\bar\theta\, W\frac{D^2}{\partial^2}W  \, ,
\label{dddd}
\eeq
which, certainly, reduces to Eq. (\ref{khorop}) but at the  price
of an explicit infrared singularity.  

The Wilsonean action, deprived of the infrared contributions by 
construction,
would not contain the term (\ref{plokhp}) or (\ref{dddd}).
 We concluded  that, if the
theory is regularized in the infrared domain, the gauge term in 
the effective action is renormalized {\em only} at  one loop. 
The Wilsonean coupling gets no corrections beyond one loop.

It is worth stressing that in the given context
the ``infrared contribution" has nothing to do with the distances
$\sim \Lambda^{-1}$. We  mean rather the contribution associated 
with
the virtual momenta $p$ of order of the external momentum
carried by the background field, as opposed to the ultraviolet 
contribution associated with $p\sim M_{\rm UV}$. The external 
momentum can be as large as we want, it plays the role of the 
running parameter in the renormalization-group evolution.

The nonrenormalization theorem above
is akin to the one we had proven for the instanton measure,
where a natural infrared regularization is provided by the instanton 
size $\rho$.
In fact, the proof is quite similar; it follows from the analysis
of the supergraphs of Fig. 4 in the chiral background field.
This time, unlike the instanton analysis,
one assumes the  background field to be weak (and chiral). One 
expands in $W$, keeping only the quadratic  terms in  $W$.

As a result, the gauge term
in the Wilsonean action is renormalized only at one
loop,$\,$\footnote{Equations (\ref{sgccr}) and (\ref{sganom}) refer to 
supersymmetric gluodynamics. Their extensions valid in the general 
theory
with  arbitrary matter  are presented below.}
\beq
\frac{1}{g^2}= \frac{1}{g^2_0}- \frac{3T_G}{8\pi^2}\,\ln\frac{M_{\rm 
UV}}{\mu}\, .
\label{sgccr}
\eeq
Correspondingly, the superanomaly  written in operator form is 
also one-loop,
\begin{equation}
\bar{D}^{\dot\alpha}{\cal J}_{\alpha\dot\alpha} =
-\frac{T_G}{8\pi^2}\,D_\alpha {\rm Tr}\,W^2\, .
\label{sganom}
\end{equation}
Thus, it was not the anomaly in the Adler-Bardeen current that had 
to  be reinterpreted, but, rather, the anomaly in the trace of the 
energy-momentum tensor. 

The idea that  the anomaly in the trace of the energy-momentum
tensor is proportional to the $\beta$ function was so deeply rooted 
that the simple step reflected in Eq. (\ref{sganom}) was painfully 
difficult  to  make. As I mentioned, it took us  years of long  
debates. In view of the importance of the issue it is worth rephrasing
the result somewhat differently.
The operator anomaly in the trace of the energy-momentum
tensor is
\beq
\theta^\mu_{\,\mu} = - \frac{3T_G}{8\pi^2}\, G_{\mu\nu}^a 
G^{a\mu\nu}\, .
\eeq
This expression  is exact.
The higher order terms  in $g^2$ on the right-hand side
 can appear only at the stage of taking 
the
matrix element of the operator $G^2$ in the given background
field.$\,$\footnote{Let me note in passing that the  anomaly
in $\theta^\mu_{\,\mu}$ is {\em not} proportional to the full $\beta$ 
function
in (nonsupersymmetric) QCD either.  The question arises  at two loops. 
To get the
anomaly in operator form one must carefully single out (and
discard) the infrared contribution. Surprisingly, this has never been 
done, in
spite of  the mature age of QCD. Why? At two loops
virtually all calculations with gluons are done in dimensional 
regularization, which does not allow one to easily separate the infrared 
part.  Therefore,  the
answer is unknown till the present day.}

In the general case of the gauge theory
with matter $\{\Phi_i\}$ and arbitrary superpotential ${\cal W}$ the  
superanomaly relation takes the form
\begin{eqnarray}
\bar{D}^{\dot\alpha}J_{\alpha\dot\alpha}&\!\!= &\!\!\frac{2}{3} 
D_{\alpha}\left\{
\left[ 3{\cal W } - \sum_i \Phi_i\, \frac{\partial{\cal W }}{\partial 
\Phi_i}
\right] \right.
 \nonumber\\[0.2cm]
&\!\!- &\!\!
\left.
\left[ \frac{3T_G- \sum_i T(R_i)}{16\pi^2}\,{\rm Tr}\,W^2 + 
\frac{1}{8}\sum_i\gamma_i 
\bar{D}^2 (\bar{\Phi}_{i}\, e^{V} {\Phi}_{i}) \right]
\right\}\, .
\label{geom}
\end{eqnarray}
The first line comes from a classical calculation, the second line presents 
the anomaly. 
This result was obtained
almost 15 years ago;$\,$\cite{SVanom} I will comment
on its derivation momentarily. The general
superanomaly relation (\ref{geom}) was confirmed recently from  an
unexpected side.  It turns out that the expression in the braces
determines$\,$\cite{DS,CS} the central charge in the central extension 
of the ${\cal N}=1$ superalgebra. The anomalous term in the central 
charge is obtained by combining Eq. (\ref{geom}) with the Konishi 
anomaly,$\,$\cite{Konishi}
\begin{equation}
  \bar{D}^2\, (\bar{\Phi}_i e^{V} {\Phi}_i) = 
4 \,{\Phi}_i \frac{\partial{\cal W }}{\partial {\Phi}_i} +
\frac{T(R_i)}{2\pi^2}\,{\rm Tr}\, W^2\, .
\label{ka1}
\end{equation} 
  Then, the coefficient in front of ${\rm Tr}\,W^2$ in the central 
charge comes out proportional to
$T_G- \sum_i T(R_i)$. In supersymmetric  QCD it vanishes provided
that the number of colors $N$ is equal to  the number of flavors $N_f$.
The vanishing of the anomalous term  in the central charge is an
indispensable feature of Seiberg's  solution$\,$\cite{Sei2} of the 
$N_f=N$ theory. Unfortunately, I do not have time to dwell on details of
this intriguing theme,  a comprehensive explanation can be found in the
review  paper$\,$\cite{SVrr} which just appeared.

If the coefficient of ${\rm Tr}\, W^2$ in the superanomaly (\ref{geom})
is purely one-loop, where does the
multiloop $\beta$ function come from? This is a legitimate question.
To answer it, let us have a closer look at the right-hand side
of the superanomaly relation. Assume  that the superpotential ${\cal 
W}$ vanishes (this assumption is not crucial, it just facilitates the task).

In the second line of Eq. (\ref{geom}) we deal with a quantum 
operator. The full $\beta$ function emerges in passing to
the matrix element of this operator in the given ($c$-number) 
background 
field. It is convenient to carry out the transition in two stages.
First, eliminate $\bar{D}^2\, (\bar{\Phi}_i e^{V} {\Phi}_i) $
in favor of ${\rm Tr}\, W^2$ by virtue of the Konishi
formula. This is still an operator relation. It is seen that at this stage the 
numerator of the NSVZ $\beta$ function is recovered. At the second 
stage we convert the quantum operator ${\rm Tr}\, W^2$ into
${\rm Tr}\, W^2_{\rm bkgr}$ where the
subscript bkgr means background. This conversion gives rise
to the denominator of the NSVZ $\beta$ function.

\section*{The Wilsonean action}

The solution of the superanomaly problem is intertwined
with another subtle question which was put forward$\,$\cite{SVanom}
in 1986 -- the distinction between the Wilsonean and canonic actions.
Surprisingly, before our work people did not realize that
this distinction existed and was instrumental in understanding the
analytic properties of supersymmetric theories in the gauge coupling 
constant.
So far my definition of the Wilsonean action and its canonic counterpart
was operational and rather vague. The distinction
 is best illustrated
in the theories with matter, where its origin is absolutely
transparent. Assume we have a  supersymmetric gauge theory with 
 arbitrary
matter $\{ \Phi_i\}$. Assume that at the ultraviolet cut off
the Lagrangian is
\begin{eqnarray}
{\cal L}&=& \left\{ \frac{1}{4g^2}\, 
\int \! {\rm d}^2 \theta\, \mbox{Tr}\,W^2 +{\rm H.c.}\right\} +
\frac{1}{4}\sum_i\,  \int \! {\rm d}^2\theta {\rm d}^2\bar\theta\, 
\bar 
\Phi_i
e^V\Phi_i\nonumber
\\[0.2cm]  &+& \left\{\frac{1}{2} \,\int \! {\rm d}^2 \theta\,  {\cal 
W}(\Phi_i) 
+{\rm 
H.c.}\right\} 
  \,,
\end{eqnarray}
where ${\cal W}$ is the superpotential. After evolving down to 
$\mu$,
the effective  Lagrangian becomes
\begin{eqnarray}
{\cal L}_{\rm W}&\!\! = \!\! &\left\{ \frac{1}{4}\left[ \frac{1}{g^2}-
\frac{3T_G-\sum T(R_i)}{8\pi^2}\,\ln\frac{M_{\rm UV}}{\mu}\right]\, 
\int \! {\rm d}^2 \theta\, \mbox{Tr}\,W^2 +{\rm H.c.}\right\} 
\nonumber\\[0.2cm]
 &\!\! +\!\!  &
\frac{1}{4}\sum_i\, Z_i \left(\frac{M_{\rm UV}}{\mu}\right) \int \! 
{\rm
d}^2\theta {\rm d}^2\bar\theta\,
\bar
\Phi_i e^V\Phi_i \nonumber\\[0.2cm]
  &\!\! +\!\!  & \left\{\frac{1}{2} \,\int \! {\rm d}^2 \theta\,  {\cal
W}(\Phi_i)  +{\rm
H.c.}\right\} \, ,
\label{dopdop}
\end{eqnarray}
where $Z_i$ stands for the $Z$ factor of the matter field $\Phi_i$.
Equation~ (\ref{dopdop}) presents the Wilsonean
effective action. It immediately entails, in turn, Eq. (\ref{geom}).
A remarkable feature of supersymmetric
theories is the complexification of the gauge coupling, see the second 
line in Eq.~(\ref{susyym}),
\beq
\frac{1}{g^2}\longrightarrow \frac{1}{g^2}- 
i\,\frac{\vartheta}{8\pi^2}\, .
\label{complexcc}
\eeq
The real part of $g^{-2}$ is the conventional coupling constant with 
which one deals, say, in perturbation theory. The imaginary part is 
related to  the
vacuum angle. The Wilsonean action preserves the complex 
structure in $g^{-2}$, due to the fact that
the renormalization of ${\rm Tr}\,W^2$ is exhausted by one loop.
It goes without saying that the
complex structure, wherever it appears, is a very precious theoretical 
asset. As we will see shortly, in the canonic action the property of 
analyticity is lost.

The kinetic 
terms in Eq. (\ref{dopdop}) are normalized noncanonically. We would 
like to pass to a  $c$-number functional (sometimes called the
generator of the one-particle irreducible vertices $\Gamma$).
Note that our $\Gamma$ is identical to what is called
the ``canonic effective action'' in the current literature. 
Calculation of $\Gamma$ is equivalent to the canonic normalization of 
the kinetic terms.

It is best to start 
from the matter fields. Again, we will assume that ${\cal W} = 0$.
Passing to the canonically normalized matter, naively we 
would say that the factors $Z_i$ have no impact whatsoever and can be 
omitted.  In fact, they do have an impact.
The easiest way to detect the impact of the $Z_i$ factors
is to expand in $Z_i -1$, assuming that $Z_i$'s are close to unity.
The linear term of expansion is unambiguously fixed by 
 the Konishi anomaly (\ref{ka1}). 
Once we realized that the linear term in $Z_i -1$ emerged in the canonic 
effective action, it was not difficult to figure out the full answer.
Elimination of the $Z_i$ factors of the matter fields in the canonic
action requires that
 $T(R_i) \ln M_{\rm UV}$ 
in front of Tr$W^2$ be replaced by $T(R_i) \ln (M_{\rm 
UV}/Z_i)$,
\beq
T(R_i) \ln M_{\rm UV} \longrightarrow T(R_i) \ln \frac{M_{\rm 
UV}}{Z_i}\, .
\label{iaddm}
\eeq 

To complete the transition to the canonic
effective action one must analyze the same effect in the
gauge sector.  Denote  by $g_c$ the canonic gauge coupling.
It is the canonic gauge coupling that is routinely used in all
perturbative calculations. Usually the subscript $c$ is omitted.
I will keep it for a while to emphasize the distinction between the
Wilsonean and canonic couplings. 

One observes that $\mbox{Re}\, g^{-2}_c$   is nothing but the $Z$ factor 
of the  gauge  fields and gauginos.
 In the transition  to the canonically normalized gauge 
kinetic term, the replacement to be done is
\beq
\ln M_{\rm UV}\, \longrightarrow \, \ln \frac{M_{\rm 
UV}}{Z^{1/3}} = \ln \frac{M_{\rm 
UV}}{[{\rm Re}\, (g^{-2}_c)]^{1/3}}\,.
\label{ggrf}
\eeq
The power of $Z$ is different from that for the chiral 
superfields  (one third  {\em versus} unity, cf. Eq. (\ref{iaddm}))
because of the different  spin weights,
but the essence is the same. This is explained in detail in 
Ref.~47. Its main thrust was on the infrared manifestation of the 
anomaly.  

Every anomaly has two faces -- ultraviolet and infrared --
and can be revealed in both ways. The fact of equivalence is 
elementary and was discussed in the literature many times (see 
e.g. the review$\,$\cite{MSanomrev}). For instance, the chiral anomaly
in supersymmetric theories can be obtained as a pole in the axial 
current, or, alternatively,
 as an  ultraviolet anomaly in  the 
measure.$\,$\cite{Konishi2} The same is true for the anomaly associated 
with the rescaling of the gluon/gluino fields displayed in
Eq. (\ref{ggrf}). 
Recently it  was
rederived$\,$\cite{AHM,AHM1} from the ultraviolet side, 
from the noninvariance of the measure. This is analogous to the
Konishi--Shizuya derivation of the Konishi anomaly.

(Let me parenthetically note that the absence of the
explicit separation of the ultraviolet and infrared contributions
led the authors$\,$\cite{AHM,AHM1} to a misinterpretation
of the anomaly supermultiplet. In fact, they introduce
a ``second" energy-momentum tensor. As I have just discussed,
in the case at hand
DR works  as the infrared rather than the ultraviolet regulator.
In addition, I would like to warn that Refs.~42, 63
introduce some  confusion in the nomenclature. The coupling constant
$g_h^2$ which the authors call holomorphic is not, since it includes 
logarithms of the $Z$ factors of the matter fields. I am aware of no 
quantity which would depend on $g_h^2$ holomorphically.
In what follows I will reserve the term ``holomorphic" for the 
Wilsonean coupling, the coefficient of ${\rm Tr}\, W^2$ in the 
Wilsonean action.) 

The following relation between the Wilsonean and canonic couplings
 ensues 
\beq
\frac{1}{g^2}= \frac{1}{g^2_c}- \frac{T_G}{8\pi^2}\, \ln 
\,\mbox{Re}\,
\frac{1}{g^2_c}\, .
\label{holcan}
\eeq
Assembling all these elements together we readily find the
$\beta$ function for the canonic coupling. It
is identical to the NSVZ $\beta$ function
quoted above in connection with the instanton derivation, see 
Eq.~(\ref{nsvzbetaf}).

As was mentioned, in supersymmetric theories the gauge coupling is 
complexified, as indicated in  Eq. (\ref{complexcc}).  
The complex structure of the coefficient in front of Tr$W^2$ is 
preserved  if and only if the action  is not renormalized beyond one 
loop.  This property  is inherent to the Wilsonean action. At the same 
time, the $Z$ factors of  the fields  (including those of gluons and 
gluinos) depend on $1/g^2$ nonholomorphically (via Re$g^{-2}$).  
That is why upon  the transition to the canonical coupling
one looses the holomorphy. 
The occurrence 
of $\ln Z$, or $\ln \,\mbox{Re}\,g^{-2}_c$ for the gauge fields,
 in the transition to the canonically normalized 
effective action  was repeatedly emphasized and illustrated
in many ways in our 1986 paper.
Nonetheless, apparently, this point is difficult to understand.

Shortly after my arrival to the US in 1990 I discussed the issue
of the Wilsonean {\em versus} canonic coupling with Dan Freedman.
He told me that our presentation of this topic did not seem
clear to him, to put it mildly. I was surprised to hear that, because I
thought that everything was crystally transparent. So, I ignored his
comment. Well, life
shows that he was right and I was wrong.
This is seen from the fact that several extended commentaries
were published recently. They add no new  physical 
content in the problem, just reinterpret the 1986 results in  
different terms.
Yet, these commentaries are  perceived by many as a ``substantial 
clarification." Is this a language barrier, or a cultural difference, or 
both?
I do not know. This is not the first time I find myself in a similar 
situation. You have just heard, in Arkady's talk,
that our penguin paper, being absolutely correct,
was thought to be totally wrong for several years.
Four  referees explained to us, one after another, that it contradicts
the Glashow--Iliopoulos--Maiani cancellation. The penguin mechanism 
was accepted only after Mary K. Gaillard advocated it in
one of her review talks.

\section*{Holomorphic anomaly}
 
The coefficients  of various $F$ terms which may be present in the 
action (e.g. the matter mass terms, the inverse couplings $g^{-2}$, the 
Yukawa couplings) can be promoted to  auxiliary chiral 
superfields. The original coupling constants are then treated as the
expectation values of the auxiliary  superfields. Now, in many 
instances the subject of  analysis is itself a chiral operator, for 
example, the operator $\mbox{Tr}\, \lambda^2$.  
In these cases the outcome of
the  analysis must depend on the expectation values of the chiral 
superfields, the antichiral ones cannot enter. This means that
the chiral quantities must depend on the chiral parameters 
holomorphically.

The holomorphic dependence is an exceptionally powerful tool in 
explorations of the gauge dynamics in the strong coupling regime.
In essence, the Seiberg-Witten revolution  of 1993/94 was based on 
the power of holomorphy. I have discussed various uses of 
holomorphy which were elaborated in the
1980's, in particular, the exact determination of the
gluino condensate. In the SU(2) model, with one flavor,
$\langle\mbox{Tr}\lambda\lambda\rangle \, \propto\, \sqrt{m}$. 
The fact that the conjugate parameter $\bar m$ does not appear
in $\langle\mbox{Tr}\lambda\lambda\rangle$,
is instrumental in establishing the square root  dependence on $m$.

The inverse gauge coupling $g^{-2}$ is also a chiral parameter. Hence,
one can expect a holomorphic dependence of
$\langle\mbox{Tr}\lambda\lambda\rangle$ on $g^{-2}$ too.
Surprisingly, examination of Eq. (\ref{gc}) shows that this is not the 
case. Indeed, while $g^{-2}$ in the exponent is the complexified 
coupling constant, in the pre-exponential factor we actually deal with 
Re$\, g^{-2}$, rather than with the full complex $g^{-2}$.
 Had $g^{-2}$ appeared in the 
pre-exponential factor, the $\vartheta$ dependence of the gluino 
condensate would come out wrong.

Does this mean that something went wrong with the general argument? 
Yes and no.  There are two gauge couplings --
canonical  and the one that enters in the Wilsonean action.
The proper question to ask is, to  which coupling does the proof of  
holomorphy refer.
This question was not asked until 1991. Only then did we realized in 
full$\,$\cite{SV} that it is always the Wilsonean coupling 
that one deals with 
in the statement of holomorphy (that's where the term holomorphic
coupling comes from). If one expresses the gluino
condensate (\ref{gc}) in terms of the Wilsonean coupling
by virtue of Eq. (\ref{holcan}), one  gets
$\langle\mbox{Tr}\lambda\lambda\rangle =\mbox{Const.}\, \exp (-
4\pi^2/g^2)$. This dependence is perfectly holomorphic. 

At the same time,  holomorphy is violated for the canonical 
coupling, due to infrared singularities. This is called the holomorphic 
anomaly. The loss of holomorphy is associated with the $Z$ factors
which are nonchiral and get entangled, with necessity, as soon as we 
pass to the canonical gauge coupling. 

\begin{figure}   
\epsfxsize=9.5cm
\centerline{\epsfbox{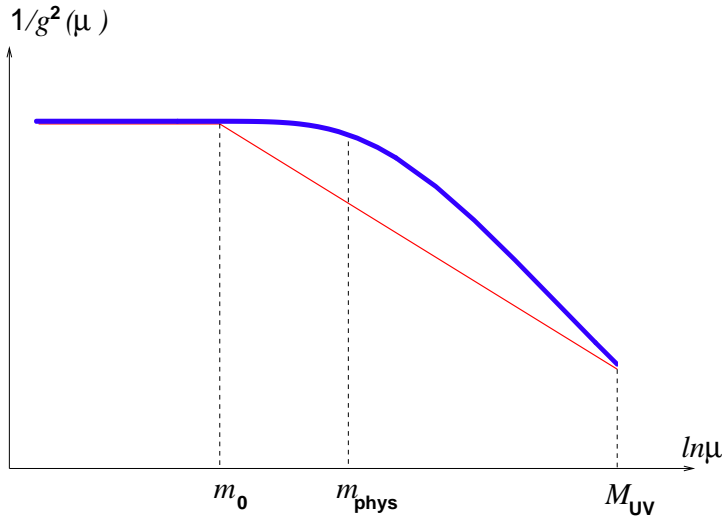}}
 \caption{Evolution of the gauge coupling in SUSY QED (schematic). 
The straight lines represent the one-loop evolution with the fake 
bare threshold $m_0$. The actual evolution is given by the smooth 
curve. The outcome at $\mu\ll m$ is the same.}
\end{figure}

The most graphic illustration of the phenomenon which I can think of
can be given in SUSY QED.  Denote the bare mass term of the electron 
$m_0$ and the bare coupling constant $g_0$.  At the ultraviolet 
cutoff $M_{\rm UV}$ the holomorphic and canonical couplings coincide. 
As one  descends from $M_{\rm UV}$ down to lower values of the 
normalization point $\mu$ they diverge. The $\beta$ function 
for the canonic coupling is multiloop, see Eq. (\ref{bfsqed}). The 
second and higher loops are entirely due to the anomalous dimension 
of the electron (selectron) field and are in one-to-one correspondence 
with the loss of holomorphy. The running $g^{-2}(\mu)$ depends on 
$g_0^{-2}$ nonholomorphically. However,  once the evolution is 
completed (i.e. at $\mu\ll m$)  and the gauge coupling freezes, the 
holomorphic dependence on  $g_0^{-2}$ and $m_0$ is restored.
In fact, one obtains the low-energy (frozen) $g^{-2}$ by using the 
one-loop (holomorphic) $\beta$ function, with a fictitious value of 
the threshold, $m_0$, 
\beq
\left.\frac{1}{g^2}\right|_{\mu\ll m} = \frac{1}{g^2_0}+ 
\frac{1}{4\pi^2}\,\ln\frac{M_{\rm UV}}{m_0}\, , 
\eeq
rather than the physical threshold, which is given by the physical 
electron (selectron) mass. The latter, in turn, is a nonholomorphic 
function of $g_0^2$. 
The nonholomorphic dependence of $m$ on $g_0^2$ combines
with the nonholomorphic part in the $\beta$ function to
cancel each other.
This is illustrated in Fig. 5 (for a more detailed 
discussion see Ref.~46).

This situation is general. When the chiral quantity measured is a 
``final product", summarizing dynamics at all scales,
it is expressible in terms of the Wilsonean coupling in a holomorphic 
way. At the same time, the snapshot {\em en~route},
at a given value of $\mu$,  captures  the canonical coupling
which carries the violation of holomorphy.$\,$\cite{SV}

This solution could have been found much earlier, in 1986.
We had all necessary elements handy, but missed the point then.
The late 1980's were an especially hard time for me personally, for 
various reasons. Life in the capital of the last world empire had 
always been like the theater of the absurd, except it was real. I could 
not 
stand it anymore, and I could not focus on physics. The explorations 
of the holomorphy issue were resumed only in 1990
when both Arkady and I moved to Minneapolis.
When our paper was essentially written,  we received a 
preprint$\,$\cite{DKL}
by Dixon, Kaplunovsky and Louis. These authors came across 
a similar holomorphic anomaly in a mass parameter, in the context of 
stringy 
calculations at one loop. They identified the reason lying
behind the anomaly as an infrared singularity due to the propagation 
of massless matter fields in the loop. It was yet another 
manifestation of the general phenomenon we had worked on. It was 
staggering to see how  the parallel lines of reasoning led to one and 
the same conclusion.  Later, Vadim  Kaplunovsky
told me that the apparent loss of holomorphy  in the string 
calculation they had  done  baffled them for quite some time, and 
he  was startled by the treatment of the problem  in our paper.

\section*{Unsolved mystery of $4/5$}

Now I have to return to the gluino condensate to fulfill several 
promises made in passing. We already know that it was obtained
by various distinct methods: at first, from the correlation function 
(\ref{tpf}) at short distances, and, later, 
$\langle\mbox{Tr}\, \lambda^2\rangle$ was obtained in the Higgs 
regime. The key element of the first derivation was  cluster 
decomposition. Instrumental in the second derivation was the
holomorphic dependence of $\langle\mbox{Tr}\, \lambda^2\rangle$
on the bare mass parameter of the auxiliary matter. Although
the functional dependence of $\langle\mbox{Tr}\, \lambda^2\rangle$
on the ultraviolet cutoff and the gauge coupling comes out the same 
in both methods, the numerical coefficients are different! (Cf.
Eqs. (8) and (12) which exhibit
a mismatch factor $\sqrt{4/5}$.) Since the
discrepancy is certainly not due to an algebraic error, something
conceptual  must have been  overlooked. The calculation
based on the Higgs regime and holomorphy seems ironclad.
The only plausible explanation
suggested so  far$\,$\cite{KScsv} was a chirally symmetric vacuum,
whose  existence in 
supersymmetric gluodynamics could have an impact on the strong 
coupling calculation.
(I leave aside explanations associated with fantastic creatures like 
tensionless strings.) Let me explain this in more detail.

The model described by the Lagrangian (\ref{susyym}) is invariant
 with respect to
the phase rotations of the gluino field (the chiral rotations).
This symmetry, valid at the classical level, is broken by the
triangle anomaly. A discrete chiral  $Z_{2N}$ symmetry
survives, however, as an exact quantum symmetry.$\,$\cite{EW} The
gluino condensate is noninvariant with respect to the  chiral 
$Z_{N}$ rotations, it
breaks (spontaneously) $Z_{2N}\to Z_2$. Consequently, if the 
gluino 
condensate develops, it can take $N$ different values which mark the
distinct chirally asymmetric vacua of the
theory.  For instance, in the SU(2) model the
condensate is double-valued, see Eq. (\ref{gc}). 

To elucidate the reason why the weak and strong coupling 
calculations
of $\langle\mbox{Tr}\,\lambda^2\rangle$ may differ,
we may invoke the hypothesis due to Amati {\em et 
al.}$\,$\cite{AMATIr}
(remember, the one which I 
was reluctant to accept in 1985, in  heated  debates 
with Amati and Veneziano,
and  appreciated only a decade  later). 
According to this hypothesis,
the strong coupling calculation of the correlation function
$\langle \mbox{Tr}\,\lambda^2(x)\,\, 
\mbox{Tr}\,\lambda^2(0)\rangle$ 
 yields, in fact,  a
result  averaged over {\em all} vacuum states of the theory. Assume
there
exists a chirally symmetric vacuum,$\,$\cite{KScsv} with
$\langle \mbox{Tr}\,\lambda^2\rangle =0$. Then, it would 
contaminate
the correlation function (\ref{tpf}), thus explaining a suppression
factor that popped out$\,$\cite{zeroSV} in  the strong coupling regime 
compared to 
the
calculation at weak 
coupling. In the latter, a large vacuum expectation value 
of the squark field picks up the vacuum state  
unambiguously -- in the SU(2) model it has to be one of two 
chirally asymmetric vacua.

There are arguments in favor   of and against  this unexpected 
chirally symmetric vacuum. An additional indication of its existence
is provided by the Veneziano-Yankielowicz effective
Lagrangian,$\,$\cite{veneziano,KScsv} and its subsequently extended 
versions.$\,$\cite{gabadadze} I must admit that
the available evidence is circumstantial, at best.

If the vacuum at $\langle \mbox{Tr}\,\lambda^2\rangle =0$
does exist  its properties must be quite exotic.
The  chirally symmetric vacuum must give zero contribution to 
Witten's 
index since the latter is fully saturated by the chirally asymmetric 
vacua.  This implies
that massless fermions are mandatory in the  
$\mbox{Tr}\, \langle\lambda^2\rangle 
=0$ phase of SUSY gluodynamics. If so, it is potentially unstable 
under
 various
deformations.  For instance, putting the system in a finite-size box
lifts the vacuum energy density from zero.$\,$\cite{KKS} 
This vacuum disappears in finite volume. This instability -- the 
tendency to 
escape under seemingly ``harmless" deformations --  may explain
why the   vacuum at 
$\langle\mbox{Tr}\lambda^2\rangle =0$ is not seen in Witten's 
$D$-brane construction.$\,$\cite{EWD}  Perhaps, this is not surprising 
at 
all. 
Indeed, there is a good deal of extrapolation in this construction, 
against which the chirally asymmetric vacua are stable
(they have no choice since they have to saturate Witten's index)
while the $\langle\mbox{Tr}\lambda^2\rangle =0$ vacuum need not 
be stable and may not  survive the space-time distortions associated 
with the $D$-brane engineering.  Neither is  it seen 
in the Seiberg-Witten solution$\,$\cite{SEIW} of ${\cal N} = 2$ SUSY
gluodynamics slightly perturbed by a small mass term for the
 matter field $m$Tr$\,\Phi^2$,  ($m\ll
\Lambda$). In this model 
the chirally symmetric state $\mbox{Tr}\, \langle \lambda^2 
\rangle= 
\langle
m\Phi^2
\rangle=0$ resembles a sphaleron: it realizes a saddle 
point in the profile of energy.
 If the chirally symmetric vacuum
develops it can happen only  at large values of $m$, i.e. $m\gg
\Lambda$. 

\section*{Conclusions}

The study of the analytic properties of  supersymmetric theories,
which began in the 1980's,
brought lavish fruits in the 1990's. The arsenal of tools based
on holomorphy expanded. The range of applications grew even more
dramatically, especially after the fundamental works of  
Seiberg and Witten in 1994. 
At the same time, the end of the road is not even in sight.
The list of profound  unanswered questions in QCD, related to 
phenomena
at large distances, is almost as large now as it was twenty years ago,
in spite of extremely impressive progress in numerous  applied 
problems.
The potential of the holomorphy-based methods in the prototype
supersymmetric gauge theories is far from being exhausted.

\section*{One last general remark}

On the last pages of the book {\em The Character of Physical Law},  
Feynman writes$\,$\cite{2} about two  alternative  scenarios of what 
can  happen to  physics ``at the very end". Either  all fundamental 
laws  will be found and we will be able to predict  everything; the  
predictions will  always be in full accord with experiment. Or it will 
turn out that new experiments will become  too  expensive or too 
complicated technically, so that we will understand  about 99.9\% of 
physical
phenomena, leaving the remaining 0.1\% of  inaccessible phenomena 
without solid  theory.  One will have to wait for a long time until  
new extremely difficult and expensive experiments are  done, so 
that the cognitive process becomes exceedingly slow and 
uninteresting. 
Feynman notes that he was very lucky to live in a time when great 
discoveries in high energy physics could be made. He compares 
his time with the discovery of America, which was discovered once 
and 
forever. This can never be repeated again.
Some theorists of my generation believe that this may well be the 
case,
the glorious days of high energy physics are over.

I do not think so. It is certainly true that the
most fundamental theory of the day,  string theory and its 
offsprings ($M$ theory, $D$ branes, {\em etc.}),  operate with the
Planck  scale which lies so far away from  the (present) human scale, 
that 
there is no  hope of carrying out  direct experimental studies. 
I do not know whether  we will
be able to advance without  direct experimental guidance,  led only 
by 
aesthetical principles.
My prime interests lie in QCD and other gauge theories. No matter 
what
happens at the Planck scale, new developments in $M$ theory and 
$D$ 
branes
give new insights to QCD practitioners. They
have already produced a strong impact on our
understanding of qualitative features of  QCD.
Let me mention, for instance,  Witten's observation of an infinite set
of vacua in QCD in the limit $N\to\infty$. This result was first 
obtained 
in 
the stringy
context,$\,$\cite{EWDD} and only later was demonstrated  in field 
theory.$\,$\cite{MSSS}
The powerful tools of supersymmetry were instrumental
in both cases. There are  good reasons to believe that more advances 
are 
about to
come. 

\section*{Acknowledgments}

It is a great honor to have been awarded the J.J. Sakurai Prize for 
Theoretical Particle Physics. For me this event has a special significance. 
I am deeply  grateful to the
American Physical Society. I was moved by warm congratulations from
many friends and colleagues, to whom I want to say thank you. 
I am grateful to my coauthors
Arkady Vainshtein, Valya Zakharov, Vitya Novikov, Misha Voloshin,
Ian Kogan,  Gia Dvali, Alex Kovner, and Boris Chibisov, with
whom I shared fun and excitement of the journey in supersymmetric 
gauge theories that started in 1981 and continues till present. 

I am grateful to A. Gorsky, A. Ritz, A. Vainshtein,  and V. Zakharov for 
useful comments in the process of conversion of notes which I had 
prepared for the talk at the Centennial Meeting of the American 
Physical Society into this  expanded written version. The 
work  was supported in part by DOE under the grant number
DE-FG02-94ER40823.

\end{document}